\documentclass{emulateapj}

\usepackage{natbib}
\usepackage{amsmath}
\usepackage{color}

\newcommand{\w}{\mathrm{w}}
\newcommand{\ej}{\mathrm{ej}}
\newcommand{\s}{\mathrm{s}}

\shorttitle{Diversity of Type IIn Supernovae}
\shortauthors{Moriya \& Maeda}

\begin{document}

\title{
CONSTRAINING PHYSICAL PROPERTIES OF
TYPE II{\MakeLowercase n} SUPERNOVAE \\
THROUGH RISE TIMES AND PEAK LUMINOSITIES
}

\def\aifa{1}
\def\kyoto{2}
\def\ipmu{3}

\author{
{Takashi J. Moriya}\altaffilmark{\aifa} and
{Keiichi Maeda}\altaffilmark{\kyoto,\ipmu}
}

\altaffiltext{\aifa}{
Argelander Institute for Astronomy, University of Bonn,
 Auf dem H\"ugel 71, D-53121 Bonn, Germany; moriyatk@astro.uni-bonn.de
}
\altaffiltext{\kyoto}{
Department of Astronomy, Kyoto University,
Kitashirakawa-Oiwake-cho, Sakyo-ku, Kyoto 606-8502, Japan
}
\altaffiltext{\ipmu}{
Kavli Institute for the Physics and Mathematics of the Universe (WPI),
Todai Institutes for Advanced Study,
University of Tokyo, Kashiwanoha 5-1-5, Kashiwa, Chiba 277-8583, Japan
}

\begin{abstract}
We investigate the diversity in the wind density, supernova ejecta
 energy, and ejecta mass in Type IIn supernovae based on
their rise times and peak luminosities.
We show that the wind density and supernova ejecta properties can be
estimated independently if both the rise time and peak luminosity 
are observed.
The peak luminosity is mostly determined by the supernova properties
and the rise time can be used to estimate the wind density.
We find that
the ejecta energy of Type IIn supernovae
needs to vary by factors of $0.2-5$ from the average
if their ejecta mass is similar.
The diversity in the observed rise times
indicates that their wind density 
varies by factors of $0.2-2$ from the average.
We show that Type IIn superluminous supernovae should have not only
large wind density but also large ejecta energy and/or small ejecta mass
to explain their large luminosities and the rise times at the same time.
We also note that
the shock breakout does not necessarily occur in the wind even if 
it is optically thick, except for the case of superluminous supernovae,
and we analyze the observational data both with and without assuming
that the shock breakout occurs in the dense wind of Type IIn supernovae.
\end{abstract}


\section{Introduction}
Type IIn supernovae (SNe IIn) are a class of SNe in which the
signatures of the interaction between the SN ejecta and the circumstellar
medium are observed \citep{schlegel1990,filippenko1997}.
The estimated circumstellar density required to explain
the observational properties is much higher than that expected
from the standard stellar evolution theory (e.g., \citealt{langer2012}).
It is generally assumed that the high circumstellar density is due to
the high mass-loss rates of the SN IIn progenitors.
The estimated mass-loss rates are typically higher than
$10^{-3}$ $M_\odot~\mathrm{yr^{-1}}$
(e.g., \citealt{kiewe2012,taddia2013,fransson2013,moriya2014}).

SNe IIn are heterogeneous.
For example, their peak luminosities spread in more than two orders of
magnitudes (e.g., \citealt{richardson2014,li2011}).
The diversities can be caused by many reasons, e.g.,
the diversities in circumstellar density, SN ejecta mass,
and SN ejecta energy.
It is important to understand which observational properties are
affected by which physical parameters.
By disentangling the origins of the observational diversities, we can constrain
the physical properties of the progenitor systems and obtain the 
better understanding of SNe IIn and their progenitors.

In this Letter, we investigate a way to disentangle the information
on the wind density, SN ejecta energy, and SN ejecta mass based on
the early light curves (LCs) of SNe IIn.
We suggest that the wind and SN properties can be constrained 
independently if both the rise time and the peak luminosity of a SN IIn
are observed.
By using the observational rise times and peak luminosities of SNe IIn
recently reported by \citet{ofek2014}, we show how diverse
the wind and SN properties should be to explain the observational
diversities of SNe IIn.

\citet{ofek2014} also performed a similar analysis by using their data
mainly focusing on the rise times when they reported the observations.
However, they did not use the peak luminosities to constrain the SN properties.
In addition, they assumed that the shock
breakout always occurs in the dense wind in SNe IIn.
The shock breakout does not necessarily occur in the wind
even if it is optically thick and makes SNe IIn.
Here, we also investigate the case in which the shock breakout does not
occur in the wind.

\section{Diffusion Time and Characteristic Luminosity}\label{section2}
We analytically estimate the diffusion time $t_d$ and the characteristic
luminosity $L_p$ of SNe IIn at $t_d$
based on the way presented by \citet{chevalier2011}.
\subsection{Diffusion Time}
The diffusion time $t_d$ is estimated as $t_d=\tau_\w \Delta R/c$,
where
$c$ is the speed of light,
$\tau_\w$ is the Thomson scattering optical depth of the wind from the
radius $R_b$ where photons start to be emitted in the wind,
$\Delta R$ is the length between $R_b$ and
the radius where the wind optical depth becomes unity
($r=R_{\tau_\w=1}$).
We assume the steady-wind density structure $\rho_\w=Dr^{-2}$.
$D$ can be expressed by using the progenitor's mass-loss rate $\dot{M}$
and the wind velocity $v_\w$ as $D=\dot{M}/(4\pi v_\w)$.
Assuming that the wind radius and $R_{\tau_\w=1}$ is much larger than $R_b$,
the diffusion time is expressed as
\begin{equation}
t_d=\frac{\tau_\w \Delta R}{c}=\frac{\kappa^2 D^2}{c R_b},\label{td}
\end{equation}
where $\kappa$ is opacity and assumed to be 0.34 $\mathrm{cm^2~g^{-1}}$
below.
$R_b$ differs depending on whether the shock breakout occurs in the wind
or not. We derive $t_d$ in the two cases separately.

\subsubsection{Shock Breakout Model}
If the shock breakout occurs in the wind, photons in the shock are
released when the following condition is satisfied
\begin{equation}
\tau_\w\simeq \frac{c}{v_\s},
\end{equation}
where $v_\s$ is the shock velocity.
We assume that the SN ejecta density structure has two density components,
$\rho_\ej\propto r^{-n}$ outside and $\rho_\ej\propto r^{-\delta}$
inside (see, e.g., \citealt{chevalier2011}) and
the SN ejecta expands homologously.
Then the radius and velocity of the shock evolve following the power-law
analytic formula with time $t$ \citep[e.g.,][]{chevalier1982,moriya2013}.
The shock breakout occurs at
\begin{equation}
t_{br}\simeq\frac{n-3}{n-2}\frac{\kappa D}{c}.
\end{equation}
Using the power-law formula presented in \citet{moriya2013}
\begin{equation}
r_\s(t)=C_1 D^{-\frac{1}{n-2}}
M_\ej^{-\frac{n-5}{2(n-2)}}
E_\ej^{\frac{n-3}{2(n-2)}}
t^{\frac{n-3}{n-2}},\label{rs}
\end{equation}
to estimate the shock radius at $t=t_{br}$ which is $R_b$,
we obtain the diffusion time $t_d$ for the case of the shock breakout in
the dense wind
\begin{equation}
t_d\simeq C_2 \kappa^{\frac{n-1}{n-2}}
D^{\frac{n}{n-2}} M_\ej^{\frac{n-5}{2(n-2)}} E_\ej^{-\frac{n-3}{2(n-2)}},\label{tdbr}
\end{equation}
where
$E_\ej$ is the kinetic energy of the SN ejecta and
$M_\ej$ is the mass of the SN ejecta (see Table \ref{tabletd}).
The constants $C_1$ and $C_2$ are shown in Appendix.

\subsubsection{No Shock Breakout Model}
If the total optical depth of the wind is smaller than $c/v_\s$
when the shock reaches the inner radius $R_i$ of the dense wind,
the shock breakout does not occur in the dense wind.
In this case, $R_b=R_i$ and the diffusion time in the wind is
\begin{equation}
t_d\simeq\frac{\kappa^2D^2}{cR_i}.\label{tdnbr}
\end{equation}
For example, when $\dot{M}=10^{-3}$ $M_\odot~\mathrm{yr^{-1}}$ and
$v_\w=100$ $\mathrm{km~s^{-1}}$, the total wind optical depth does not exceed
$c/v_\s \simeq 30$ if $R_i\gtrsim 6\times 10^{12}$ cm with the standard
$v_\s\simeq 10000\ \mathrm{km~s^{-1}}$.
It is possible that the progenitor radius is larger than
$\simeq 6\times 10^{12}$ cm $=86\ R_\odot$ and the wind does not become optically thick
enough to cause the shock breakout in it.
This dividing radius is
less than those of red supergiants (RSGs) and luminous blue variables (LBVs)
$(\sim 100\ R_\odot)$.
Even if the progenitor radius is smaller than 
$\simeq 6\times 10^{12}$ cm, it is possible that the dense wind
does not start just above the progenitor and there exist a 'void'
between the progenitor and the dense part of the wind.
In addition, if $v_\s$ is smaller, the shock breakout radius can be
smaller. For instance, if $v_\s\simeq 5000\ \mathrm{km~s^{-1}}$ which
is indicated in some SNe IIn in \citet{ofek2014},
$R_i$ needs to be smaller than $3\times 10^{12}$ cm to cause the shock breakout.
Thus, we do not assume that the shock breakout always occurs in the
dense wind and investigate the case in which the shock breakout does not occur.

\subsection{Characteristic Luminosity}
We estimate the characteristic luminosity at $t_d$ by
assuming that a fraction of
the kinetic energy in the SN ejecta shocked in $t_d$
is radiated in $t_d$.
The total available kinetic energy $E_p$ is 
\begin{equation}
E_p= \int^\infty_{r(t_d)}4\pi r^2 \frac{1}{2}\rho_\ej v_\ej^2
 dr,\label{Eavailable}
\end{equation}
where $v_\ej$ is the SN ejecta velocity.
Eq. (\ref{rs}) is used to estimate $r(t_d)$
in deriving Eq. (\ref{Eavailable}).

Assuming that a fraction $\epsilon$ of the kinetic 
energy is emitted in the diffusion time,
we obtain the characteristic luminosity $L_p$,
\begin{equation}
L_p\simeq\frac{\epsilon E_p}{t_d}.
\end{equation}

By using $t_d$ obtained in the previous section, $L_p$ can be
expressed as a function of $t_d$, $M_\ej$, and $E_\ej$
(see Table \ref{tableLp}).
When the shock breakout occurs in the wind, we get the characteristic
luminosity
\begin{equation}
L_p\simeq C_3\epsilon \kappa^{-\frac{(n-5)(n-1)}{n(n-2)}}
t_d^{\frac{n^2-10n+10}{n(n-2)}}
M_\ej^{-\frac{(4n-5)(n-5)}{2n(n-2)}}
E_\ej^{\frac{(4n-5)(n-3)}{2n(n-2)}}.\label{Lpbr}
\end{equation}
For no shock breakout case, we get
\begin{equation}
L_p\simeq C_4 \epsilon \kappa^{-\frac{n-5}{n-2}}
R_i^{\frac{n-5}{2(n-2)}}
t_d^{\frac{n-11}{2(n-2)}}
M_\mathrm{ej}^{-\frac{3(n-5)}{2(n-2)}}
E_\mathrm{ej}^{\frac{3(n-3)}{2(n-2)}}
.\label{Lpnbr}
\end{equation}
The constants $C_3$ and $C_4$ are shown in Appendix.

\subsection{Summary}
We summarize the dependence of $t_d$ and $L_p$ 
on the wind and SN properties in Tables \ref{tabletd} and \ref{tableLp}
with specific examples for typical $n$.
$t_d$ has strong dependence on the wind properties,
while $L_p$ has strong dependence on the SN properties.

Using the relations obtained in this section, the physical properties of
SN ejecta and the wind can be constrained separately as
\begin{equation}
 M_\ej^{-\frac{(4n-5)(n-5)}{2n(n-2)}}E_\ej^{\frac{(4n-3)(n-3)}{2n(n-2)}}
= C_3^{-1}\epsilon^{-1}\kappa^{\frac{(n-5)(n-1)}{n(n-2)}}L_p
t_d^{-\frac{n^2-10n+10}{n(n-2)}},\label{snbr}
\end{equation}
\begin{equation}
D=C_2^{-\frac{n-2}{n}}C_3^{-\frac{n-2}{4n-5}}
\epsilon^{-\frac{n-2}{4n-5}} \kappa^{-\frac{3(n-1)}{4n-5}}
L_p^{\frac{n-2}{4n-5}} t_d^{\frac{3(n-1)}{4n-5}},\label{Dbr}
\end{equation}
for the case with the shock breakout and
\begin{equation}
 M_\ej^{-\frac{3(n-5)}{2(n-2)}}E_\ej^{\frac{3(n-3)}{2(n-2)}}
=C_4^{-1}\epsilon^{-1}R_i^{-\frac{n-5}{2(n-2)}}\kappa^{\frac{n-5}{n-2}}
L_p t_d^{-\frac{n-11}{2(n-2)}},\label{snnbr}
\end{equation}
\begin{equation}
 D=\sqrt{\frac{cR_i t_d}{\kappa^2}},\label{Dnbr}
\end{equation}
for the case without the shock breakout.
These relations show that 
the wind and SN ejecta properties can be constrained independently
if both $t_d$ and $L_p$ are observed.
The SN ejecta properties strongly affected by $L_p$
and the wind properties by $t_d$.


\begin{deluxetable}{lcccc}
\tablecaption{Power-law dependence of $t_d$}
\tablehead{
Model & $D$ & $R_i$ & $M_\ej$ & $E_\ej$
}
\startdata
breakout (general) & $\frac{n}{n-2}$ &$0$  & $\frac{n-5}{2(n-2)}$  &
 $-\frac{n-3}{2(n-2)}$\\ 
breakout ($n=10$) & $1.25$ &$0$& $0.313$ &$-0.438$ \\ 
breakout ($n=7$)  & $1.4$ &$0$& $0.2$& $-0.4$ \\ 
no breakout& $2$&$-1$& $0$& $0$ 
\enddata
\tablecomments{Based on Eqs. (\ref{tdbr}) and (\ref{tdnbr}).}
\label{tabletd}
\end{deluxetable}

\begin{deluxetable*}{lcccc}
\tablecaption{Power-law dependence of $L_p$}
\tablehead{
Model & $R_i$  & $t_d$ & $M_\ej$ & $E_\ej$
}
\startdata
breakout (general) & $0$&$\frac{n^2-10n+10}{n(n-2)}$ &
$-\frac{(4n-5)(n-5)}{2n(n-2)}$  &
 $\frac{(4n-5)(n-3)}{2n(n-2)}$\\ 
breakout ($n=10$) & $0$&$0.125$ & $-1.09$ & $1.53$ \\ 
breakout ($n=7$)  & $0$&$-0.314$& $-0.657$& $1.31$ \\ 
no breakout (general)&$\frac{n-5}{2(n-2)}$& $\frac{n-11}{2(n-5)}$& $-\frac{3(n-5)}{2(n-2)}$& $\frac{3(n-3)}{2(n-2)}$ \\
no breakout ($n=10$)&$0.313$&$-0.0625$ & $-0.938$ & $1.31$ \\
no breakout ($n=7$)& $0.2$&$-0.4$ &$-0.6$ & $1.2$ 
\enddata
\tablecomments{Based on Eqs. (\ref{Lpbr}) and (\ref{Lpnbr}).}
\label{tableLp}
\end{deluxetable*}

\begin{figure}
 \begin{center}
  \includegraphics[width=\columnwidth]{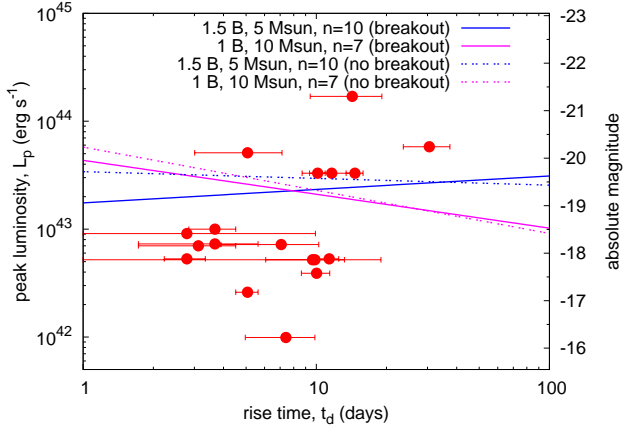} 
 \end{center}
\caption{
Rise times and peak luminosities of SNe IIn
with analytic relations between the diffusion time $t_d$
and the characteristic luminosity $L_p$.
Data points are based on Table 1 of \citet{ofek2014}.
No bolometric correction is assumed.
Characteristic luminosities $L_p$ (Eq. \ref{Lpbr} for the shock
 breakout and Eq. \ref{Lpnbr} for no shock breakout)
with two parameter sets
are plotted in the figure, ($E_\ej$/ 1 B, $M_\ej/M_\odot$, $n$)
$=(1.5,5,10)$ and $(1,10,7)$. 
}\label{projecta}
\end{figure}

\section{Diversity in Type II{\MakeLowercase n} Supernovae}
In this section, we investigate the origin of the observational
diversities in the rise times and the peak luminosities of SNe IIn
and relate these observational diversities
to the diversities in the wind density and SN ejecta
properties in SNe IIn.
We use the analytic estimates for $t_d$ and $L_p$ obtained in the
previous section for this purpose.
\citet{ofek2014} recently summarized the rise times and peak
luminosities of SNe IIn. Based on their Table 1, we plot
the rise times and peak luminosities of SNe IIn in Fig. \ref{projecta}.
The peak luminosities are based on the $R$ band data and we do not adopt
any bolometric corrections in the figure.
The bolometric corrections for SNe IIn change with time, but
are typically within 0.5 mag and small \citep{ofek2014b}.
We assume that the rise time corresponds to $t_d$ and the peak luminosity corresponds
to $L_p$.

\begin{figure*}
 \begin{center}
  \includegraphics[width=0.9\columnwidth]{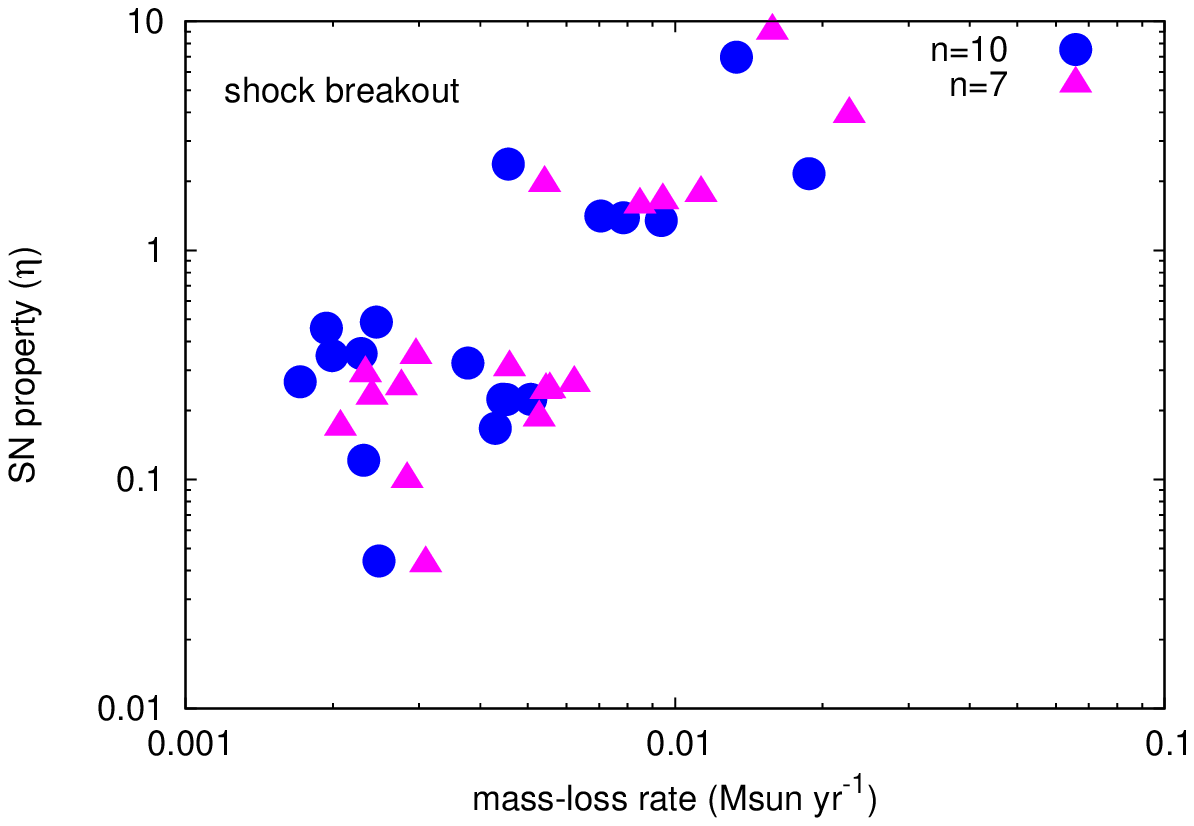} 
  \includegraphics[width=0.9\columnwidth]{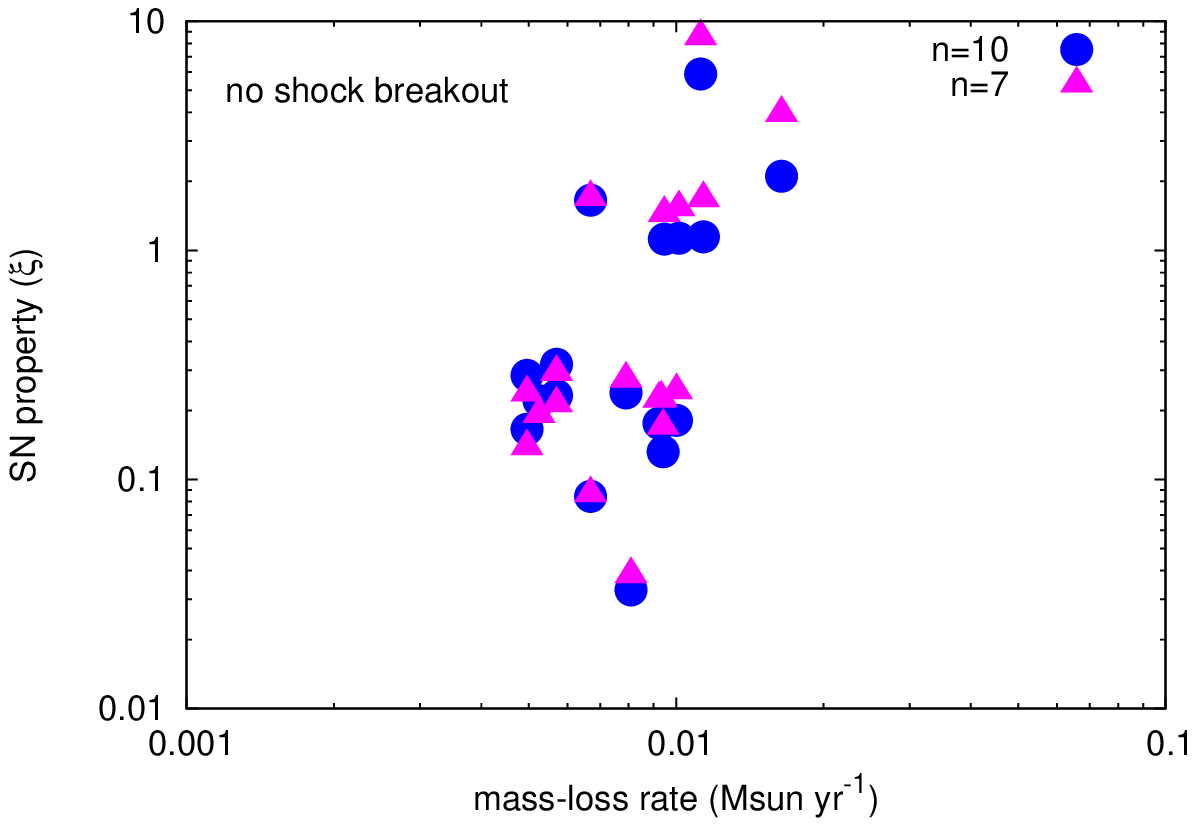} 
 \end{center}
\caption{
Diversities in the wind and SN properties in SNe IIn estimated from observed
 $t_d$ and $L_p$.
The mass-loss rates are estimated with $\dot{M}=4\pi v_\w D$ and
$v_\w=100$ $\mathrm{km~s^{-1}}$.
The SN properties are
$\eta=(M_\ej/M_{\mathrm{ej,s}})^{-1.09}(E_\ej/E_\mathrm{ej,s})^{1.53}$
 $(n=10)$ or
$\eta=(M_\ej/M_{\mathrm{ej,s}})^{-0.657}(E_\ej/E_\mathrm{ej,s})^{1.31}$ $(n=7)$
 (Eq. \ref{snbr}),
 and
$\xi=(M_\ej/M_\mathrm{ej,s})^{-0.938}(E_\ej/E_\mathrm{ej,s})^{1.31}$
 $(n=10)$ or
$\xi=(M_\ej/M_\mathrm{ej,s})^{-0.6}(E_\ej/E_\mathrm{ej,s})^{1.2}$
 $(n=7)$ (Eq. \ref{snnbr}).
In our model, $E_{\ej,\s}=1.5$ B and $M_{\ej,\s}=5$ $M_\odot$ for $n=10$
and $E_{\ej,\s}=1$ B and $M_{\ej,\s}=10$ $M_\odot$ for $n=7$.
The standard values depend on the model assumptions but the dispersion
does not. The left panel shows the parameters for the case of the shock breakout
 and the right panel shows the region for the case of no shock breakout.
}\label{disp}
\end{figure*}


\subsection{Diversity in Supernova Ejecta}
The analytic estimates for the peak luminosity (Eqs. \ref{Lpbr} and \ref{Lpnbr})
show that, if the diffusion time is known, the peak luminosity is
mostly determined by the SN properties ($M_\ej$ and $E_\ej$).
In Fig. \ref{projecta}, $L_p$ is plotted for two sets of SN ejecta
properties,
$(E_\ej/ 1\ \mathrm{B},M_\ej/M_\odot,n)=(1.5, 5, 10)$ and $(1, 10, 7)$
\footnote{1 B $=10^{51}$ erg}, for the cases with and without
the shock breakout.
The conversion efficiency $\epsilon$ is set as 0.3.
The suggested conversion efficiency ranges $\epsilon\simeq 0.1-0.5$
in literature and we choose an average value \citep[see, e.g.,][]{fransson2013}.
We set $\delta=0$ and choose two $n$ based on the previous studies
(e.g., \citealt{chevalier2011,fransson2013,matzner1999}).
The characteristic luminosities from the two parameter sets roughly correspond
to the average peak luminosity of SNe IIn in Fig. \ref{projecta}.
The exact values of $M_\ej$ and $E_\ej$ which give the average $L_p$
depend on the model assumptions like in $\epsilon$ but the diversity does not.
As we can see in Fig. \ref{projecta}, $L_p$ does not strongly depend on
$t_d$ and it is in fact mostly determined by the SN ejecta properties.
We can see in Fig. \ref{projecta} that the differences in
the observational peak luminosities are roughly within the factors of $0.1-10$ to
the analytical average lines.
If the shock breakout occurs in the wind,
this means that the diversity in the SN
properties is roughly within the following range,
\begin{equation}
0.1<
\eta\equiv
\left(\frac{M_\ej}{M_\mathrm{ej,s}}\right)^{-\frac{(4n-5)(n-5)}{2n(n-2)}}
\left(\frac{E_\ej}{E_\mathrm{ej,s}}\right)^{\frac{(4n-5)(n-3)}{2n(n-2)}}
<10,\label{sbrdiv}
\end{equation}
where $E_\mathrm{ej,s}$ and $M_\mathrm{ej,s}$ are the standard values.
In our case,
$E_\mathrm{ej,s}=1.5$ B and $M_\mathrm{ej,s}=5\ M_\odot$ for $n=10$,
and
$E_\mathrm{ej,s}=1$ B and $M_\mathrm{ej,s}=10\ M_\odot$ for $n=7$.
In Fig. \ref{disp}, we show the estimated diversity in the SN properties
of SNe IIn.
For example, if the exploding stars in SNe IIn have similar ejecta mass
($M_\ej\simeq M_\mathrm{ej,s}$),
the ejecta energy needs to be diversified 
by roughly factors of $0.2-5$ from the standard value.
If the ejecta energy is roughly the same in SNe IIn ($E_\ej\simeq E_\mathrm{ej,s}$), the ejecta mass
should be diversified by about factors of $0.1-8$ ($n=10$) or $0.03-40$
($n=7$) from the standard ejecta mass.
Most SNe IIn are found in the observed luminosity range in
Fig. \ref{projecta} \citep{richardson2014,li2011} and the
diversities in SN ejecta properties estimated here are presumed
to exist generally in SNe IIn.

So far, we used the shock breakout model to discuss the diversity.
The total wind optical depth estimated from Eq. (\ref{tdnbr}) exceeds
$\simeq30$ 
if $t_d\gtrsim3.5$ days with $R_i=10^{13}$ cm $=140\ R_\odot$
and the shock breakout
may not occur in some SNe IIn shown in Fig. \ref{projecta},
assuming the typical SN shock velocity of
10000 $\mathrm{km~s^{-1}}$.
The progenitor radius can be larger than 140 $R_\odot$ if the progenitor
is a RSG or a LBV.
Alternatively, if the dense wind is detached and
$R_i=10^{14}$ cm, for example, the shock breakout only occurs in the SNe IIn
with $t_d\gtrsim35$ days. The detachment can occur in SN IIn progenitors
if they have variable mass-loss rates.
We can see from Fig. \ref{projecta} that most SNe IIn have $t_d$
which is less than 35 days.
Assuming $R_i=10^{14}$ cm, the observational diversity indicates
\begin{equation}
0.1<
\xi\equiv
\left(\frac{M_\ej}{M_\mathrm{ej,s}}\right)^{-\frac{3(n-5)}{2(n-2)}}
\left(\frac{E_\ej}{E_\mathrm{ej,s}}\right)^{\frac{3(n-3)}{2(n-2)}}
<10.
\label{nsbrdiv}
\end{equation}
The expected diversity does not differ much from that expected from the
shock breakout model (Fig. \ref{disp}).
However, $L_p$ also depends on $R_i$ in this case.
The difference in $R_i$ by a factor 10 can make the difference in the
luminosity by a factor about 2 (Eq. \ref{Lpnbr}).

Whether the shock breakout occurs in the wind also strongly depends on
the shock velocity.
If the shock velocity is 5000 $\mathrm{km~s^{-1}}$,
the shock breakout occurs in SNe IIn with $t_d>14$ days
even in the wind $R_i=10^{13}$ cm $=140\ R_\odot$, which is compatible
with RSG and LBV radii.
Then, both SNe IIn with and without
the shock breakout may commonly exist (Fig. \ref{projecta}).
\citet{ofek2014} assumed that the shock breakout always occurs in the
dense wind of SNe IIn and they tried to constrain $v_\s$
by using the relation $\tau_\w\simeq c/v_\s$.
However, it does not necessarily occur in every SN IIn.
Photon diffusion in the wind
without the shock breakout may occur commonly in SNe IIn.

Fig. \ref{disp} indicates that there may exist two separate
populations in the ejecta properties, since SNe IIn do not exist
at $\eta\simeq1$ nor $\xi\simeq1$. However, the number of the observations
is still small and this remains to be investigated.

\subsection{Diversity in Wind}
If the shock breakout occurs in the dense wind in SNe IIn, 
the diffusion time $t_d$ depends both on the wind properties and the SN
ejecta properties (Eq. \ref{tdbr}) but
it is more sensitive to the wind density.
The wind density can be estimated with Eq. (\ref{Dbr}) by $t_d$ and $L_p$.
Fig. \ref{disp} shows the mass-loss rates of SN IIn progenitors
obtained by the estimated wind density ($\dot{M}=4\pi v_\w D$).
We find that the wind density in SNe IIn differs by roughly factors
of $0.2-2$ from the average when the shock breakout occurs
for the standard sets of $M_\ej$ and $E_\ej$.
The estimated mass-loss rates in Fig. \ref{disp}
ranges $\sim 10^{-3}-\sim10^{-2}$ $M_\odot~\mathrm{yr^{-1}}$ and they
are consistent with those estimated
in the previous SN IIn studies (e.g., \citealt{fox2011,kiewe2012,taddia2013,moriya2014}).

If the shock breakout does not occur in the wind,
the wind density can be estimated solely from $t_d$ with Eq. (\ref{Dnbr}).
Since $t_d$ in Fig. \ref{projecta} are roughly
between 1 day and 30 days, the corresponding
diversity in the wind density is by factors of $0.3-1.7$,
assuming a constant $R_i$ and the average $t_d$ of 15 days.
If $R_i=10^{14}$ cm, we obtain
$D=1.5\times 10^{15}$ $\mathrm{g~cm^{-1}}$ for $t_d=1$ day and
$D=8.2\times 10^{15}$ $\mathrm{g~cm^{-1}}$ for $t_d=30$ days.
Fig. \ref{projecta} shows that most $t_d$ is between 1 day and 30
days in SNe IIn
and the wind density is presumed to differ by factors of $0.3-1.7$
if $R_i$ is constant.
If the wind velocity $v_\w$ is 100 $\mathrm{km~s^{-1}}$,
the corresponding mass-loss rates of the progenitors are
$2.9\times 10^{-3}$ $M_\odot~\mathrm{yr^{-1}}$ and
$1.6\times 10^{-2}$ $M_\odot~\mathrm{yr^{-1}}$, respectively (Fig. \ref{disp}).

\section{Superluminous Supernovae}
The peak magnitudes of superluminous SNe (SLSNe)
are brighter than $-21$ mag or roughly
$10^{44}$ $\mathrm{erg~s^{-1}}$ \citep{gal-yam2012}.
\citet{quimby2013} constructed the pseudo-bolometric LCs of SLSNe.
The rise times of SLSNe IIn are typically larger than 40 days.
This means that SLSNe IIn have both large $t_d$ and $L_p$.
The large diffusion time indicates that the wind is generally dense enough
to cause the shock breakout in SLSNe as is suggested by previous works
(e.g., \citealt{chevalier2011}).
The peak luminosities are typically more than about one order of magnitude larger than
our standard $L_p$ in Fig. \ref{projecta}.
We have shown that the peak luminosity does not strongly
depend on the wind properties in the shock breakout model and it is
mostly determined by the SN ejecta properties.
This means that, if the SN ejecta mass of SLSNe is similar to other
SNe IIn, their SN kinetic energy needs to be higher by more than a
factor of 5 to explain the huge luminosities (Eq. \ref{sbrdiv}).
Alternatively, the SN ejecta mass can be smaller by a factor of
less than 0.1 $(n=10)$ or 0.03 $(n=7)$ if their SN ejecta energy
is similar to the standard SNe IIn.
The total emitted energy just by radiation in SLSNe IIn is
typically more than $10^{51}$ erg and it is likely that the SN energy
is higher than usual SNe.

We show that the large peak luminosities in SLSNe suggest large $E_\ej$
and/or small $M_\ej$. However, looking at Table \ref{tabletd},
we find that larger $E_\ej$ and smaller $M_\ej$ both make $t_d$ smaller.
However, $t_d$ in SLSNe IIn is much larger than those of SNe IIn.
To make large $t_d$ with large $E_\ej$ and/or small $M_\ej$,
the wind density must be very large.
This indicates that the extremely large explosion energy (and/or
the extremely small ejecta mass) as well as the extremely
dense wind is required 
to explain both the large diffusion times and luminosities of SLSNe.
Energetic explosions (and/or explosions with very small mass)
need to be somehow accompanied by the formation
of the dense wind.
Detailed modeling of SLSNe also indicates
the necessity of high explosion energy in the extremely dense wind
(e.g., \citealt{ginzburg2012,moriya2013b,chatzopoulos2013}).

\section{Conclusions}
We have investigated the diversities in rise times and peak
luminosities in SNe IIn and related them to the diversities in the wind
and SN properties.
We have shown that the peak luminosities are mostly affected by the SN
properties. 
The rise times which we relate to the diffusion time $t_d$ in
the wind can be used to estimate the wind properties individually.
We also note that the shock breakout does not necessarily occur in the
wind, especially if the progenitors are RSGs or LBVs, and we investigate
the models with and without the shock breakout.

The expected diversity in SN ejecta properties estimated from the
diversity in the SN IIn peak luminosities
is shown in Fig. \ref{disp}.
If the SN ejecta mass does not
differ much in SNe IIn, the diversity in the SN ejecta energy is by factors
of $0.2-5$ from the average. If the SN ejecta energy is similar in SNe IIn,
the diversity in the SN ejecta mass is expected to be by factors
of $0.1-8$ $(n=10)$ or $0.03-40$ $(n=7)$ from the average.
The expected diversity does not strongly differ if we assume that the shock
breakout occurs in the wind or not.

The diversity in the wind density can be estimated with the diversity in
the rise times (Fig. \ref{disp}). 
If the shock breakout occurs in the wind, the expected diversity
in the wind density for SNe IIn with similar peak luminosities
is by factors of $0.2-2$ from the average.
If the shock breakout does not occur, the diversity is factors of $0.3-1.7$
from the average.

SLSNe IIn show both the large peak luminosities and the large rise times.
We suggest that both the high wind density and the high explosion energy
and/or small ejecta mass are required to explain the properties of the SLSNe.
The large rise times indicate that the shock breakout occurs in the wind
in the SLSNe. The large peak luminosities indicate that the explosion energy
is very large and/or the ejecta mass is very small. However,
the large explosion energy and/or small ejecta mass make the diffusion time
smaller. Thus, the large wind density is required to have the large
rise times.
Putting together, 
not only the larger wind density but also the larger SN energy and/or
the smaller SN ejecta mass than typical SNe IIn are required
to have the large peak luminosities and large rise times at the same
time as observed in SLSNe.

\begin{acknowledgments}
TJM is supported by JSPS Postdoctoral Fellowships for Research Abroad
 (26\textperiodcentered51).
KM acknowledges financial supports by Grant-in-Aid for Scientific
 Research for Young Scientists (23740141, 26800100) and WPI Initiative,
 MEXT, Japan.
\end{acknowledgments}

\appendix
\section{Constants}
Constants which appear in the main text are
\begin{equation}
C_1=\left[
\frac{1}{2\pi(n-4)(n-3)(n-\delta)}
\frac{\left[2(5-\delta)(n-5)\right]^{\frac{n-3}{2}}}{
\left[(3-\delta)(n-3)\right]^{\frac{n-5}{2}}}
\right]^{\frac{1}{n-2}},
\end{equation}
\begin{equation}
C_2=c^{-\frac{1}{n-2}}
\left[
2\pi(n-4)(n-3)(n-\delta)
\frac{[(3-\delta)(n-3)]^{\frac{n-5}{2}}}
{[2(5-\delta)(n-5)]^{\frac{n-3}{2}}}
\right]^{\frac{1}{n-2}}
\left(\frac{n-2}{n-3}\right)^{\frac{n-3}{n-2}}
,
\end{equation}
\begin{equation}
C_3=
\frac{2\pi}{n-5}c^{\frac{n-5}{n(n-2)}}
\left[\frac{1}{4\pi(n-\delta)}
\frac{\left[2(5-\delta)(n-5)\right]^{\frac{n-3}{2}}}
{\left[(3-\delta)(n-3)\right]^{\frac{n-5}{2}}}\right]^{\frac{4n-5}{n(n-2)}} 
\left[\frac{(n-4)(n-3)}{2}\right]^{\frac{(n-1)(n-5)}{n(n-2)}}
\left(\frac{n-3}{n-2}\right)^{\frac{(n-5)(n-3)}{n(n-2)}}, 
\end{equation}
and
\begin{equation}
C_4=\frac{2\pi}{n-5}c^{\frac{n-5}{2(n-2)}}
\left[\frac{1}{4\pi(n-\delta)}
\frac{\left[2(5-\delta)(n-5)\right]^{\frac{n-3}{2}}}
{\left[(3-\delta)(n-3)\right]^{\frac{n-5}{2}}}\right]^{\frac{3}{n-2}} 
\left[\frac{(n-4)(n-3)}{2}\right]^{\frac{n-5}{n-2}}.
\end{equation}

\bibliographystyle{apj}

\begin{thebibliography}{}

\bibitem[Chatzopoulos et al.(2013)]{chatzopoulos2013} Chatzopoulos, E., 
Wheeler, J.~C., Vinko, J., Horvath, Z.~L., \& Nagy, A.\ 2013, \apj, 773, 76 

\bibitem[Chevalier(1982)]{chevalier1982} Chevalier, R.~A.\ 1982, 
\apj, 258, 790 

\bibitem[Chevalier 
\& Irwin(2011)]{chevalier2011} Chevalier, R.~A., \& Irwin, C.~M.\ 2011, \apjl, 729, L6 

\bibitem[Filippenko(1997)]{filippenko1997} Filippenko, A.~V.\ 1997, \araa, 35, 309 

\bibitem[Fox et al.(2011)]{fox2011} Fox, O.~D., Chevalier, 
R.~A., Skrutskie, M.~F., et al.\ 2011, \apj, 741, 7 

\bibitem[Fransson et al.(2013)]{fransson2013} Fransson, C., Ergon, 
M., Challis, P.~J., et al.\ 2013, arXiv:1312.6617 

\bibitem[Gal-Yam(2012)]{gal-yam2012} Gal-Yam, A.\ 2012, Science, 
337, 927

\bibitem[Ginzburg 
\& Balberg(2012)]{ginzburg2012} Ginzburg, S., \& Balberg, S.\ 2012, \apj, 757, 178

\bibitem[Kiewe et al.(2012)]{kiewe2012} Kiewe, M., Gal-Yam, A., 
Arcavi, I., et al.\ 2012, \apj, 744, 10 

\bibitem[Langer(2012)]{langer2012} Langer, N.\ 2012, \araa, 50, 107 

\bibitem[Li et al.(2011)]{li2011} Li, W., Leaman, J., 
Chornock, R., et al.\ 2011, \mnras, 412, 1441

\bibitem[Matzner 
\& McKee(1999)]{matzner1999} Matzner, C.~D., \& McKee, C.~F.\ 1999, \apj, 510, 379 

\bibitem[Moriya et al.(2014)]{moriya2014} Moriya, T.~J., Maeda, 
K., Taddia, F., et al.\ 2014, \mnras, 439, 2917 

\bibitem[Moriya et al.(2013a)]{moriya2013b} Moriya, T.~J., 
Blinnikov, S.~I., Tominaga, N., et al.\ 2013a, \mnras, 428, 1020 

\bibitem[Moriya et al.(2013b)]{moriya2013} Moriya, T.~J., Maeda, 
K., Taddia, F., et al.\ 2013b, \mnras, 435, 1520 

\bibitem[Ofek et al.(2014a)]{ofek2014} Ofek, E.~O., Arcavi, I., 
Tal, D., et al.\ 2014a, \apj, 788, 154

\bibitem[Ofek et al.(2014b)]{ofek2014b} Ofek, E.~O., Zoglauer, A., 
Boggs, S.~E., et al.\ 2014b, \apj, 781, 42 

\bibitem[Quimby et al.(2013)]{quimby2013} Quimby, R.~M., Yuan, F., 
Akerlof, C., \& Wheeler, J.~C.\ 2013, \mnras, 431, 912 

\bibitem[Richardson et al.(2014)]{richardson2014} Richardson, D., 
Jenkins, R.~L., III, Wright, J., \& Maddox, L.\ 2014, \aj, 147, 118

\bibitem[Schlegel(1990)]{schlegel1990} Schlegel, E.~M.\ 1990, 
\mnras, 244, 269 

\bibitem[Taddia et 
al.(2013)]{taddia2013} Taddia, F., Stritzinger, M.~D., Sollerman, J., et al.\ 2013, \aap, 555, A10 

\end{thebibliography}

\end{document}